\documentclass{article}

\usepackage{arxiv}

\usepackage[utf8]{inputenc} % allow utf-8 input
\usepackage[T1]{fontenc}    % use 8-bit T1 fonts
\usepackage{hyperref}       % hyperlinks
\usepackage{url}            % simple URL typesetting
\usepackage{booktabs}       % professional-quality tables
\usepackage{amsfonts}       % blackboard math symbols
\usepackage{nicefrac}       % compact symbols for 1/2, etc.
\usepackage{microtype}      % microtypography
\usepackage{lipsum}
\usepackage{graphicx}

\usepackage{amssymb}
\usepackage{multirow}
\usepackage{multicol}
\usepackage{float}
\usepackage{url}

\usepackage{bm}
\usepackage{mathenv}
\usepackage{amsmath}

\graphicspath{ {./images/} }

\title{Can the earth be flat ? A physical oceanographer's perspective}

\author{
 Charly de Marez \\
  Université de Bretagne Occidentale\\
  Laboratoire d'Océanographie Physique et Spatiale\\
  IUEM\\
  Rue Dumont D'Urville \\
  \texttt{charly.demarez@univ-brest.fr} \\
   \And
 Mathieu Le Corre \\
  Université de Bretagne Occidentale\\
  Laboratoire d'Océanographie Physique et Spatiale\\
  IUEM\\
  Rue Dumont D'Urville \\
  \texttt{mathieu.lecorre@univ-brest.fr} \\
}

\begin{document}
\maketitle
\begin{abstract}
According to a recent survey, 2\% of the U.S. population is convinced that the earth is flat. 
This idea is heralded by members of the Flat Earth Society, and promulgated through Internet forums and other public media channels.
Children and young students are easy targets.
As a science teacher confronted by these ideas, it can be a challenge to firmly confront their fundamentally flawed foundation while still remaining compassionate towards those who espouse such beliefs.
While we do not purport to have the answer for such difficult situations, as two scientists studying ocean physics, we attempt to lend a helping hand to the interested instructor.
Here, we use the most advanced computational tools of the physical oceanography community to show that the properties of the ocean that we observe from ships, satellites, and autonomous observing platforms should not exist if the earth were flat. 
In particular, using \textit{the first} realistic simulation of the ocean on a flat Earth, we show that the North Atlantic gyre, appears entirely different on a flat Earth. 
For members of the physical oceanography community, the results presented here are obvious; however, for the Flat Earth Society community, it may present an intellectual challenge.
\end{abstract}

\section{Introduction}

For the past 10 years, the idea that the earth may be flat has experienced a revival. While concrete numbers are arduous to find, rough estimates suggest that only 84\% of the U.S. population agreed with "I have always believed the world is round" (\url{https://bit.ly/2wePEmI}). Moreover, according to this survey, 2\% of the pool was deeply convinced that the earth is flat. Even if one have to be cautious with these unpublished results, it is not rare to find ---if we know where to seek--- people confident about the fact that the earth is flat, even if their scientific expertise is low. 
Supporters of the flat Earth theory are convinced that the governments and the scientific community is lying to the public by spreading false data and information. 
This idea is heralded by members of the Flat Earth Society (FES).
Children and young students are easy targets. 
As a teacher, it is difficult to be confronted to these ideas because they may be supported by students' close family, friends, or favorite artists. Following the call of the American Association of Physics Teacher \cite{noauthor_calling_2019}, it is timely to give scientific proofs, "in a calm and respectful way, to show that someone other than the science deniers" are listening to the questions of doubting people.

If the earth were flat, the dynamics of the ocean would be deeply different. 
Indeed, the large-scale circulation relies on physical mechanisms which are directly related to the sphericity of the earth. 
In this study, we propose to use state-of-the-art computational tools to compare the ocean dynamics on (1) the spherical Earth and (2) the flat Earth. 
This allows a proof, \textit{by contradiction}, that the actual properties of the ocean should not exist if the earth were flat. 
We conduct this study by using some of the most advanced tools used by the physical oceanography community today: high resolution (${\rm \Delta x\sim 6\,km}$) basin-scale realistic simulations, as well as satellite and large \textit{in situ} observation databases. 
To our knowledge, no one has tried to model the flat Earth. 
We thus present the first (and hopefully the last) realistic simulation of the ocean on a flat Earth. 
For most of the members of our Earth and planetary sciences community, the results presented here are obvious. 
Despite this, we estimate that this may have a pedagogic purpose, since we illustrate textbook-concepts phenomena in a realistic context. 
Also, our approach may provide high school and university-level science teachers with critical tools to confront and compassionately question FES views in the classroom.
Coincidentally, our presentation may help students to understand that a flat Earth is not compatible with what is actually occurring in the oceans. 
To tackle this enormous task, we have tried to conduct our study in a way that it follows as much as possible the beliefs and data espoused by the FES.

\subsection*{Model the flat flat Earth}

Understanding how the global ocean circulation might appear on a flat Earth is conceptually challenging, and we thus limit our study to the North Atlantic. Indeed, as presented by the FES on their website: "The Flat Earth is laid out like a North-Azimuthal projection. The North Pole is at the center while Antarctica is at the rim. The continents are spread out around the North Pole." 
The comparison between the oceanic currents on a flat Earth or a spherical one is made difficult because they have different horizontal extensions. 
Also, as the primary source of energy of the ocean is the atmosphere (through solar radiations and winds), one needs to consider atmospheric data to impulse energy at the surface of our simulated oceans. Looking at flat Earth official maps, one observes that the North Atlantic closely resemble the maps of the spherical Earth that others utilize (see for instance \url{https://bit.ly/2m4XuJN}). 
This would not be the case if one have examined for instance the North Pacific, or any basin in the southern hemisphere. 
This allows us, without loss of belief, to consider the North Atlantic as our study domain for both cases (spherical and flat Earth scenarios), utilizing atmospheric forcing and geography frequently encountered in this region.

We now discuss the circulation in the North Atlantic as might be found on a spherical or flat Earth. Here, we focus on the shape and intensity of the Gulf Stream, which is a strong boundary current found on the eastern coast of the United States. This latter current system is intense, flowing northward along the U.S. continental shelf from the Strait of Florida to Cape Hatteras near 40$^\circ$N, where it leaves the coast (see \textit{e.g} \cite{stommel1958gulf}, \cite{fuglister_gulf_1963}, and references therein). Studying the occurrence of this current is convenient because (1) its signature can clearly be seen at the surface of the ocean, (2) it is present throughout the whole year with little variations, and (3) its presence is accepted by the FES, such that its use is permitted within our arguments. 

Our discussion is organized as follows. In section 2, we present a set of equations that allows us to describe the ocean in a dynamical manner; in these equations, we approximate the earth as spherical and flat Earth, depending on the particular case examined. In section 3, we present our method, which includes a description of the numerical simulations and datasets used. In section 4, we compare the outputs of the two simulations with observations, and particularly the location and intensity of the Gulf Stream in order to help answer the question, "Is the earth flat?". In section 5, we discuss the simulation outputs, highlighting the physical mechanisms that modify the position and intensity of the Gulf Stream. Finally, we conclude in section 6.

\section{Governing equations for the ocean dynamics}

The dynamics of a fluid in an inertial frame of reference is described by the Navier-Stokes equations \cite{landau1987fluid}:

\begin{equation}
    \rho\frac{d\,\textbf{u}_i}{dt}=-\boldsymbol{\nabla}P+\mu \Delta \textbf{u}_i+\textbf{F}_{ext},
\label{NS}
\end{equation}
where $\rho$ is the density of the fluid (determined by an equation of state), $P$ its pressure, $\mu$ its viscosity, $\textbf{F}_{ext}$ any external forces applied to the fluid, and $\frac{d\,\textbf{u}_i}{dt}$ the acceleration of the fluid in the inertial frame.
The earth can be considered as an inertial frame of reference. Indeed, its rotation around the sun is slow (in comparison to the time scale of the fluid movements we consider, typically greater than one week), as is the rotation of the sun within the galaxy. In the flat Earth model, the earth is the center of the universe, and gravity does not exist. However, the earth is moving upward with a constant acceleration called "Universal Acceleration". By construction, this model reproduces an earth in an inertial frame of reference, with a constant force acting on all objects moving at its surface, as the gravity does. 

The acceleration of a rotating fluid can be expressed in the inertial frame of reference such as:

\begin{equation}
    \frac{d\,\textbf{u}_i}{dt}=\frac{d\,\textbf{u}}{dt}+\underbrace{2\boldsymbol{\Omega}\times \textbf{u}}_{\rm Coriolis\, force}+\overbrace{\boldsymbol{\Omega}\times(\boldsymbol{\Omega}\times \textbf{x}_i)}^{\rm Centrifugal\, force}.
\end{equation}
$\textbf{u}=(u,v,w)$ is the velocity of the fluid in the rotating frame of reference, $\boldsymbol{\Omega}$ is the rotation vector expressed in the inertial frame, and $\textbf{x}_i$ is the position of the fluid in the inertial frame. The centrifugal force is small, and can be considered as a small deviation of the gravity (or equivalent) force. It is of $O(1)$\% compared to $g=9.81\,{\rm m\,s^{-2}}$. The Coriolis force is a fictitious force that comes from the change of coordinates and cannot be neglected.
Thus, the study of a fluid at the surface of a rotating body is done by using the Navier-Stokes equation in a rotating frame of reference,

\begin{equation}
    \rho\frac{d\,\textbf{u}}{dt}=-\boldsymbol{\nabla}P+\mu \Delta \textbf{u}+\textbf{F}_{ext}-2\boldsymbol{\Omega}\times \textbf{u}.
\label{NSrot}
\end{equation}

At this point it is worth mentioning two points. (1) In many flat Earth models, the earth is not rotating. The Coriolis force thus does not act on planetary fluids. (2) The FES questions the existence of the Coriolis force (at a fundamental level), by stating that this is an "effect" which has never been observed experimentally (see \url{https://wiki.tfes.org/Coriolis_Effect}). If we consider this two points as the truth, all Newtonian mechanics and Geophysical Fluid Dynamics collapse. Here, we thus assume that Earth is rotating and that the Coriolis force does exist.

\begin{figure}
\centering
\includegraphics[width=12cm]{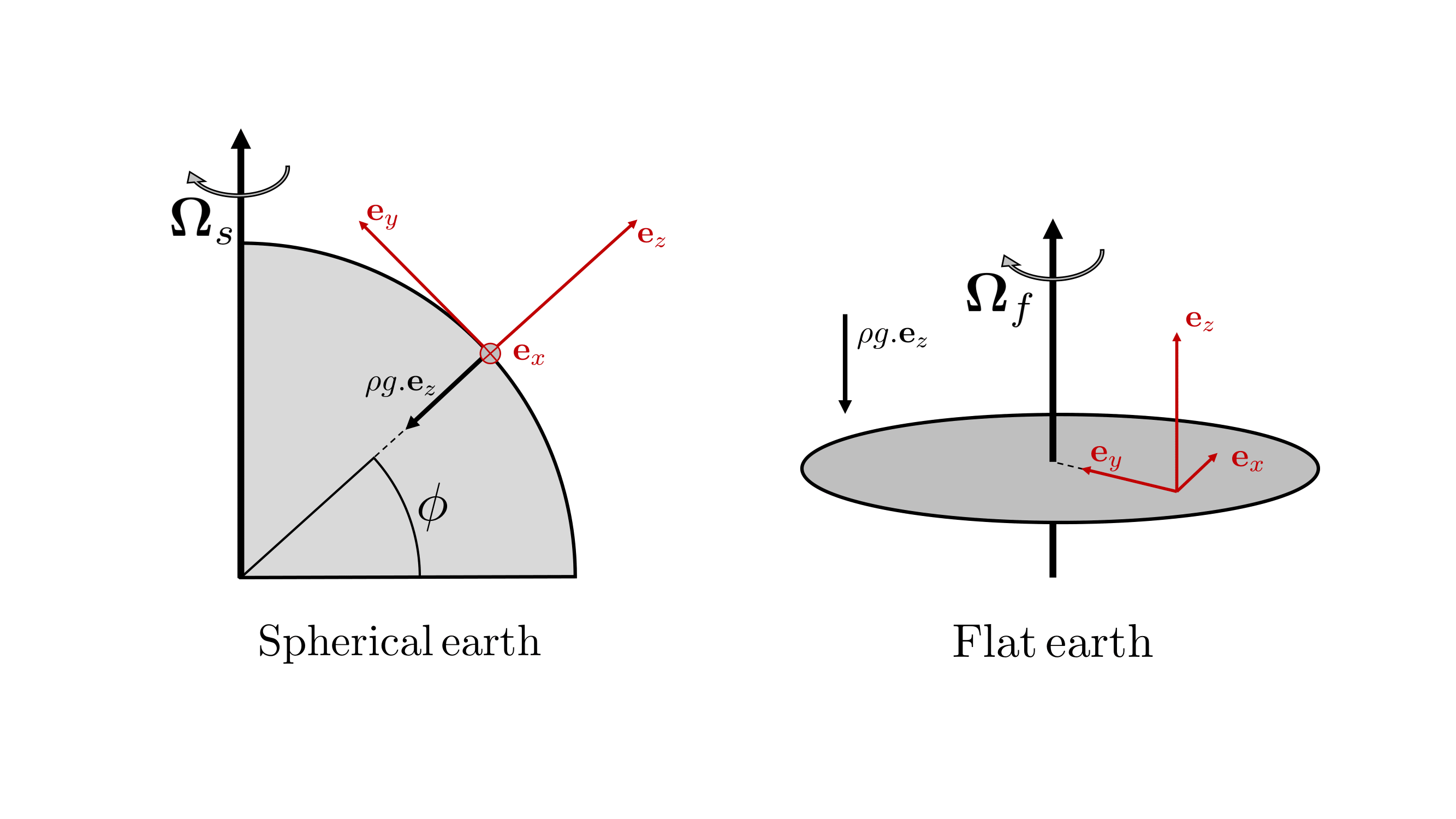}
\caption{Frame of reference for the spherical Earth (left) and the flat Earth (right). In each panel, $(\textbf{e}_x,\textbf{e}_y,\textbf{e}_z)$ are the axis of the rotating frame of reference, in which the rotation vectors are projected.}
\label{scheme}
\end{figure}

To derive equation (\ref{NSrot}), there is no need to assume any position for the rotating frame of reference. It can thus be placed on either a sphere or a disk. 
The only difference between the two cases is the projection of $\boldsymbol{\Omega}$ in the rotating frame. Considering the 2 frames of reference $(\textbf{e}_x,\textbf{e}_y,\textbf{e}_z)$ shown in Fig. \ref{scheme}, we obtain $\boldsymbol{\Omega}_s=(0,\Omega_e\,\cos\,\phi,\Omega_e\,\sin\,\phi)$ for a spherical Earth, and $\boldsymbol{\Omega}_f=(0,0,\Omega_{fe})$ for a flat Earth. $\Omega_e$ is the rotation rate of the spherical Earth, $\phi$ the latitude on the spherical Earth, and $\Omega_{fe}$ the rotation rate of the flat Earth. Applying the cross product to compute the form of the Coriolis force leads to a set of equations for the two cases. To keep equations simple, we discard viscosity and external forcings in equations. Also, the gravity (or equivalent) force is defined such as $\textbf{F}_{ext}=\rho g \,\textbf{e}_z$. This gives

\begin{equation}
  \left\{
      \begin{aligned}
        &\frac{d}{dt}u-fv+f^*w=-\frac{1}{\rho}\partial_x P  \\
        &\frac{d}{dt}v+fu=-\frac{1}{\rho}\partial_y P\\
        &\frac{d}{dt}w-f^*u=-\frac{1}{\rho}\partial_z P-g\\
      \end{aligned}
    \right.
\end{equation}
for a spherical Earth, with $f=2\Omega\sin\phi$ and $f^*=2\Omega\cos\phi$, respectively the traditional and non traditional components of the Coriolis frequency. They represent the projection of $\boldsymbol{\Omega}$ on the normal and meridional (\textit{i.e} North-South) directions of the local plane.

For the flat Earth, equations are

\begin{equation}
  \left\{
      \begin{aligned}
        &\frac{d}{dt}u-\hat{f}v=-\frac{1}{\rho}\partial_x P  \\
        &\frac{d}{dt}v+\hat{f}u=-\frac{1}{\rho}\partial_y P\\
        &\frac{d}{dt}w=-\frac{1}{\rho}\partial_z P-g\\
      \end{aligned}
    \right.
\end{equation}
with $\hat{f}=2\Omega_{fe}$.
Because the equations governing the acceleration of fluid parcels on a flat Earth neglect the meridional component of Coriolis ---the so-called traditional approximation---, we have elected to do the same for the case of the spherical Earth ($f^*=0$). This is also justified on account of the distance from the Equator and pronounced stratification in our domain.

Conceptually, the difference between a flat Earth and a spherical one can be summarized as follows: \textit{on a flat Earth there is no planetary $\beta$-effect}. Performing a Taylor expansion of the Coriolis frequency around a mean latitude $\phi_0$ on a sphere gives 

\begin{equation}
    f=f_0+\beta y +O[(\phi-\phi_0)^2],
    \label{beta}
\end{equation}
with $y$ the meridional direction, $f_0=2\Omega_e\sin \phi_0$, and $\beta=2\Omega_e\cos \phi_0/R_e$ with $R_e$ the radius of the spherical Earth. 

On the other hand, on a flat Earth, the Coriolis frequency will not vary:

\begin{equation}
\hat{f}=\hat{f}_0={\rm constant},
\end{equation}
with $\hat{f}_0=2\Omega_{fe}$. 
That is, the difference between the spherical and flat Earth is comprised in the existence of $\beta$.

To \textit{turn off} the impact of Earth's spherical nature on the dynamics, one need only set the Coriolis parameter, $f$, to a constant value. We will consider in the following the hydrostatic primitive equations to describe the planetary fluid \cite{vallis2017atmospheric}. This is a common simplification of Navier-Stokes equations, widely used in physical oceanography, which relies on the Boussinesq and the hydrostatic assumptions. This means, respectively, that (1) density is taken constant except in the gravity term, and (2) vertical acceleration of the fluid is considered small compared to the vertical pressure gradient. This gives (again discarding viscosity and external forcing terms for simplicity)

\begin{equation}
  \left\{
      \begin{aligned}
        &\frac{d}{dt}u-fv=-\frac{1}{\rho_0}\partial_x P  \\
        &\frac{d}{dt}v+fu=-\frac{1}{\rho_0}\partial_y P\\
        &\partial_z P=-\rho g,\\
      \end{aligned}
    \right.
\label{prim}
\end{equation}
with 

\begin{equation}
  \left\{
      \begin{aligned}
        &f=2\Omega_e\sin\phi \rightarrow {\rm \, spherical\, Earth}  \\
        &f=\hat{f}_0 \rightarrow{}{\rm \, flat\, Earth}.\\
        \end{aligned}
    \right.
\end{equation}

For slowly varying phenomena and large scale dynamics, the pressure gradients is in equilibrium with the Coriolis force term (in equation (\ref{prim})), this is the so-called \textit{geostrophic balance}. If we introduce a typical timescale and a horizontal lengthscale of fluid movements $T$ and $L$, this is true for $T\gg1$ day and $L>$ 100 km. Assuming that the pressure at the surface is linked to the sea surface height (hereafter SSH in the text, and $\eta$ in equations) by $\eta=\rho_s g P({\rm z=0})$, this reads

\begin{equation}
  \left\{
      \begin{aligned}
        &-fv_s=-g\partial_x \eta  \\
        &+fu_s=-g\partial_y \eta,\\
      \end{aligned}
    \right.
\label{eg}
\end{equation}
where the subscript $A_s$ designates the surface value of A ---\textit{i.e.} $(u_s,v_s)$ are the horizontal components of the surface velocity of the fluid. These (purely diagnostic) equations are useful because they allow us to describe the surface ocean dynamics simply by measuring the SSH. At large horizontal scales, this can presently be done using satellites. For example, a positive (negative) SSH anomaly corresponds to anticyclonic (cyclonic) surface currents. In the Northern hemisphere, anticyclonic currents rotate clockwise, whereas cyclonic currents rotate counter-clockwise. At lateral scales of $O(10-100)$ km, these rotating currents are refereed to as anticyclonic \textit{mesoscale eddies}. We refer the reader to \cite{chelton_global_2007,chelton_global_2011} for an example of study where mesoscale eddies are described from their SSH signature. At basin scales, we refer to such rotating currents as \textit{gyres}.
In the following, we use the SSH as well as the norm of the surface velocity, $|U|=\sqrt{u_s^2+v_s^2}$, as diagnostic quantities to describe the large scale currents.

\section{Simulating and observing the ocean}

We performed realistic simulations of the North Atlantic using the Coastal and Regional Ocean COmmunity model (CROCO, \cite{shchepetkin_regional_2005}). This model solves the hydrostatic primitive equations, \textit{i.e} the set of equations (\ref{prim}), in which the surface forcing and the bottom friction is added. The density is computed using the full equation of state for seawater \cite{shchepetkin_accurate_2011}. The horizontal advection terms for tracers and momentum are discretized with third-order upwind advection schemes (UP3), see \textit{e.g.} \cite{klein_upper_2008} for a further description. This parameterization implies implicit dissipation and it damps dispersive errors.
The horizontal resolution is ${\rm \Delta x\sim 6\,km}$ such that mesoscale eddies are reasonably well resolved. Simulations have 50 vertical levels following the topography. 
The calculations are performed in a vertically stretched coordinate such that the resolution near the surface and bottom are higher than within the ocean interior. The bathymetry is constructed from the SRTM30 PLUS dataset \cite{becker_global_2009}. The simulations are initialized, and forced at boundaries with the SODA dataset \cite{carton_reanalysis_2008}. At the surface, the forcing is taken from the daily ERA-INTERIM dataset \cite{dee_era-interim_2011}. The model parameterization are the same as the one described in \cite{lecorre_rockal_2019} and \cite{lecorre_vort_2019}. We also refer the reader to the latter references for a full validation of the numerical settings. 

We run two simulations. In the first one, we consider the Coriolis frequency $f$ of a spherical Earth, which varies with latitude. This simulation thus reproduces the ocean dynamics on a spherical Earth, including the $\beta$-effect. This simulation is called BETA hereafter in the text. We run a second simulation considering that the earth is flat, thus having a constant Coriolis frequency $f=\hat{f}_0$. This simulation is called FLAT hereafter in the text. The choice of $\hat{f}_0$ is tricky because it determines the mesoscale dynamics (10-100 km) through the Rossby radius of deformation \cite{chelton_geographical_1998}. Choosing a value of $f_0$ thus sets the eddy scale and also influences wave propagation speeds. We choose to set $\hat{f}_0=10^{-4}\,s^{-1}$ since we want to study in particular the mid-latitudes dynamics. 
%This choice is also justified by the fact that the analyses of phenomena and distances made in Flat Earth 'textbooks' surprisingly focus over U.S. and Europe. In these places, the Coriolis frequency is roughly equals to $10^{-4}\,s^{-1}$.

Simulations are run from January 1, 2000 to March 3, 2008. We consider a spin-up period of 2 years to allow the large-scale circulation to be at equilibrium. In the following, the time period we consider is thus from January 1, 2002 to March 3, 2008 ($\sim$5 year period). Averages of SSH and $|U|$ are made over the entire simulations. Averages of sea surface temperature (SST) during the winter are made for the months of December, January, and February.

We also compute the meridional heat transport of the Gulf Stream during the winter, defined by

\begin{equation}
    {\rm heat \,transport}=\int_{-1000 {\rm m}}^{0}dz\,\int_{80^\circ {\rm W} }^{60^\circ {\rm W}}dx\,\rho_0 c_p\langle v_s T \rangle_t,
\end{equation}
where the integral is performed vertically and zonaly (\textit{i.e} along the East-West direction ---$x$ is the zonal coordinate), T is the temperature, $\rho_0=1030\,{\rm kg\,m^{-3}}$ is the mean density, $c_p=3850\,{\rm J\,kg^{-1}\,K^{-1}}$ is the specific heat capacity, and $\langle . \rangle_t$ is a wintertime average over the whole period. It thus gives an estimation of the heat transport (in Watts) varying in latitude. We do the calculation between 30$^\circ$N and 40$^\circ$N. The domain considered for this calculation is shown in Fig. \ref{sst} (see black dashed rectangle).

We compare the simulation outputs with \textit{in situ} and satellite data. Surface currents are derived from the NOAA 15 m-drogued drifters dataset. For further information on these data see \cite{lumpkin_lagrangian_2007} and \cite{laurindo_improved_2017}. Values of horizontal velocities are averaged in bins of size $0.5^\circ \times 0.5^\circ$ for the period 2002-2008. The median number of $(u_s,v_s)$ data per bin is 1250, with about 5000 data per bin in the Gulf Stream region. The mean SSH is estimated by averaging the absolute dynamic topography of the AVISO dataset supplied on a daily mercator grid for the period 2002-2008. 
%It is a satellite measurement. 
%It's not a satellite measurement. Well, some of it is because it is 
%adt = sla + mdt
%but mdt is made up of a number of different products entirely independent of satellite altimetry, including drifters, etc. At larger scales, the mdt is made up of goce/grace gravity data, but at smaller scales, it becomes dominated by drifter measurements.
This product was produced and distributed by the Copernicus Marine and Environment Monitoring Service (CMEMS) (\url{http://www.marine.copernicus.eu}).

\section{Is the earth flat ?}

\begin{figure}
\centering
\includegraphics[width=16cm]{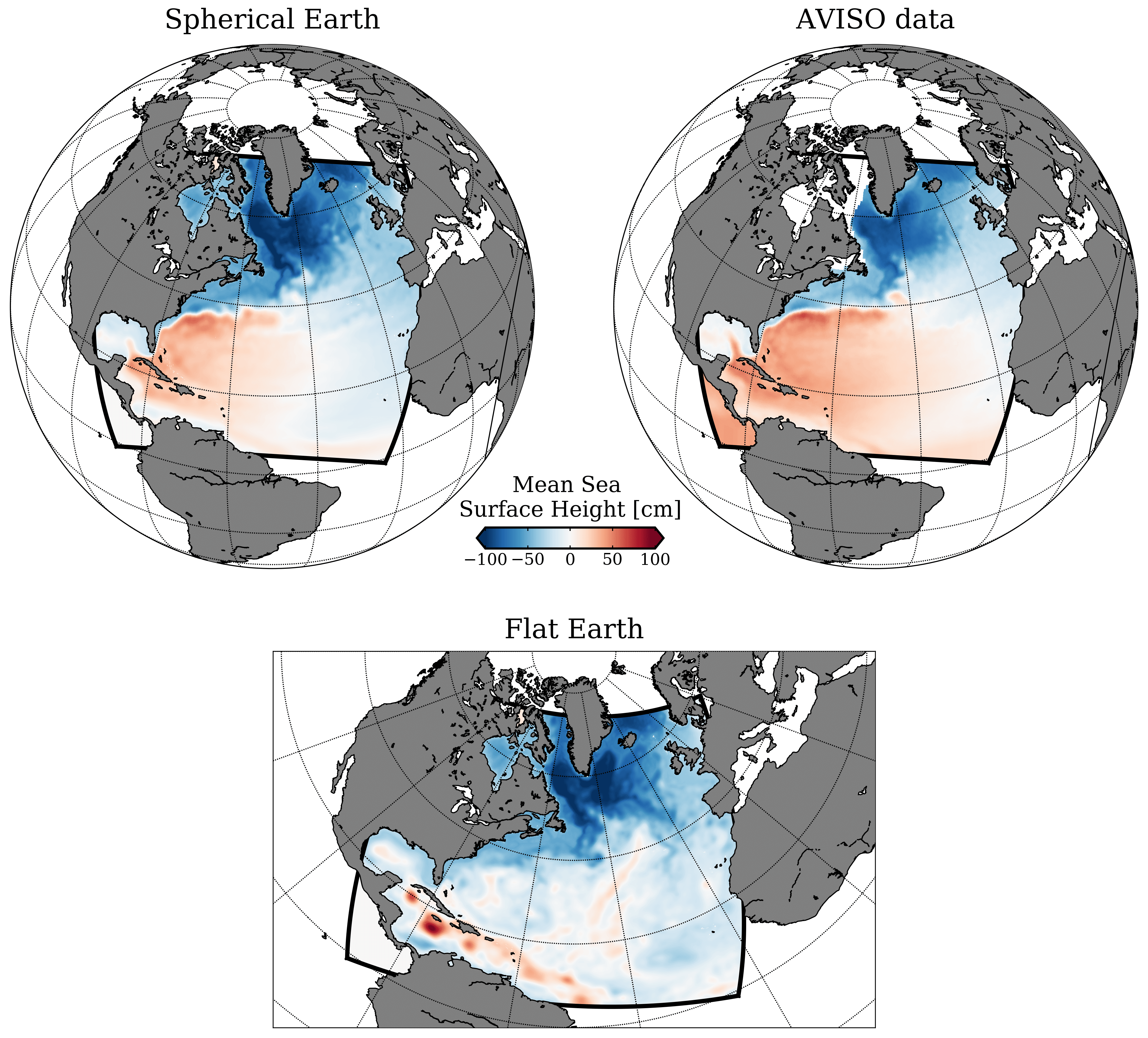}
\caption{Mean SSH in the BETA simulation (top left), the FLAT simulation (bottom), and satellite observations (top right). Averages are done over a five year period after a 2-year spin up. The BETA simulation outputs and satellite observations are projected on a sphere, while the FLAT simulation outputs are presented on a North-Azimuthal projection (following Fat Earth Society maps).}
\label{ssh}
\end{figure}

The outputs from the BETA run shows a very good agreement with satellite data, as a clear zonal dipolar structure is seen in the SSH (Fig. \ref{ssh}). This reflects the signature of the anticyclonic subtropical gyre of the North Atlantic \cite{talley2011descriptive}. In FLAT, this circulation does not set up. Also, the near-equator circulation appears to be dominated by eddies. This is not in accordance with satellite data, and is due to the fact that $f$ does not have the correct value in this region. Indeed, on a sphere, $f\rightarrow 0$ at the equator, the geostrophic equilibrium (\ref{eg}) therefore does not stand, and no coherent eddies should be seen. 

\begin{figure}
\centering
\includegraphics[width=16cm]{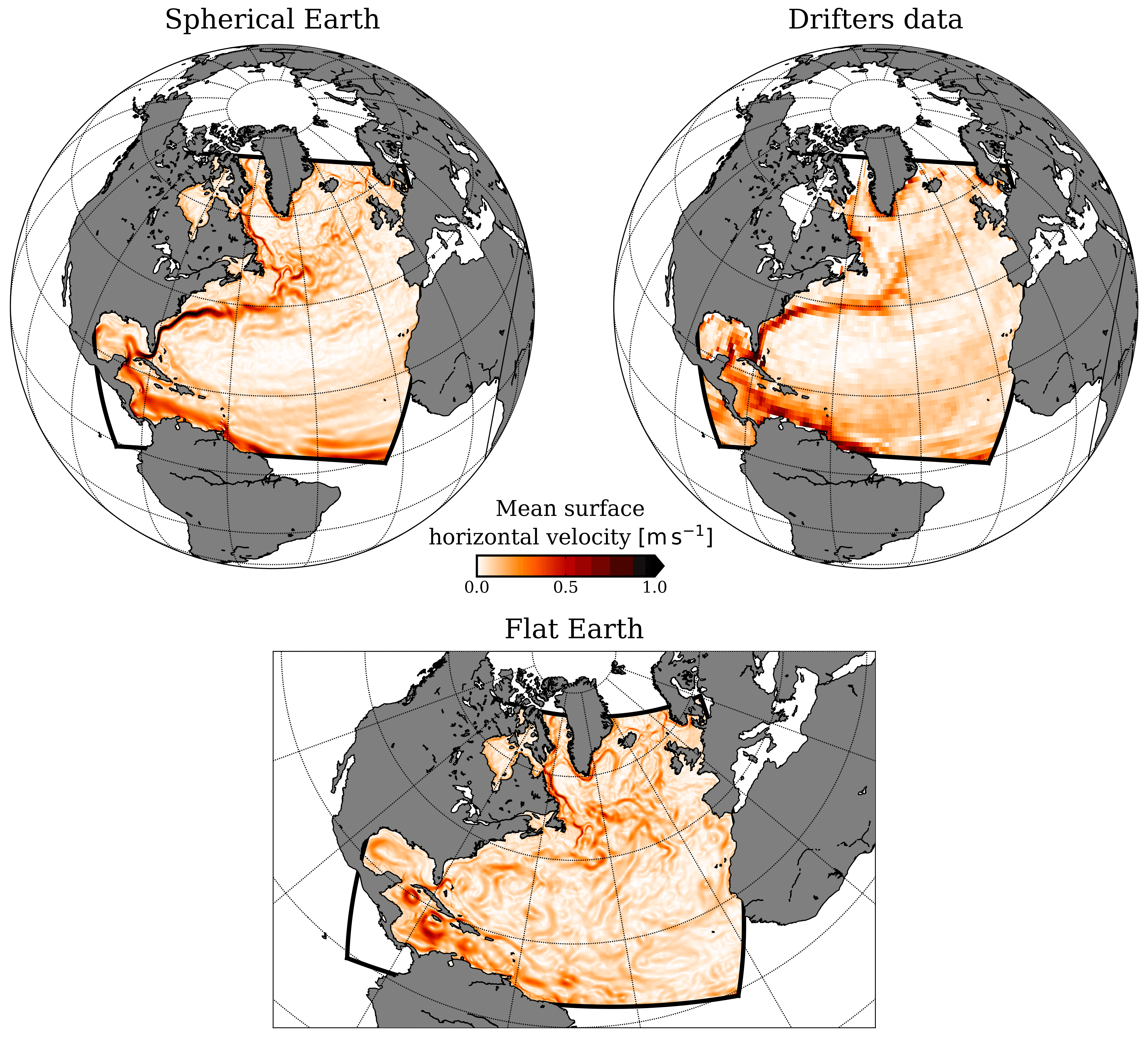}
\caption{Mean surface horizontal velocity $|U|$ in the BETA simulation (top left), the FLAT simulation (bottom), and the drifters dataset (top right). Averages are done over a five year period after a 2-year spin up. The BETA simulation outputs and drifter measurements are projected on a sphere, while the FLAT simulation outputs are presented on a North-Azimuthal projection (following Fat Earth Society maps).}
\label{current}
\end{figure}

The western part of the subtropical gyre is the Gulf Stream. Its signature is clearly seen in both the SSH (Fig. \ref{ssh}) and the surface currents (Fig. \ref{current}) of the BETA simulation. This signature is in good agreement with drifter measured surface velocities. The intensity and the width of the Gulf Stream are well reproduced near 30$^\circ$N, off the coasts of Florida, with surface velocities reaching $2\,{\rm m\,s^{-1}}$. In the FLAT simulation, the Gulf Stream signature is not seen. Furthermore, the retroflection of the Loop Current in the Gulf of Mexico \cite{hurlburt_numerical_1980}, the production and the western propagation of Loop Current rings in the Gulf of Mexico \cite{sturges_frequency_2000}, Meddies at the Soutwest of Portugal \cite{carton_meddy_2010-1}, and Gulf Stream rings \cite{robinson_gulf_1983} are well reproduced in the BETA simulation but not in FLAT (not shown).

\begin{figure}
\centering
\includegraphics[width=16cm]{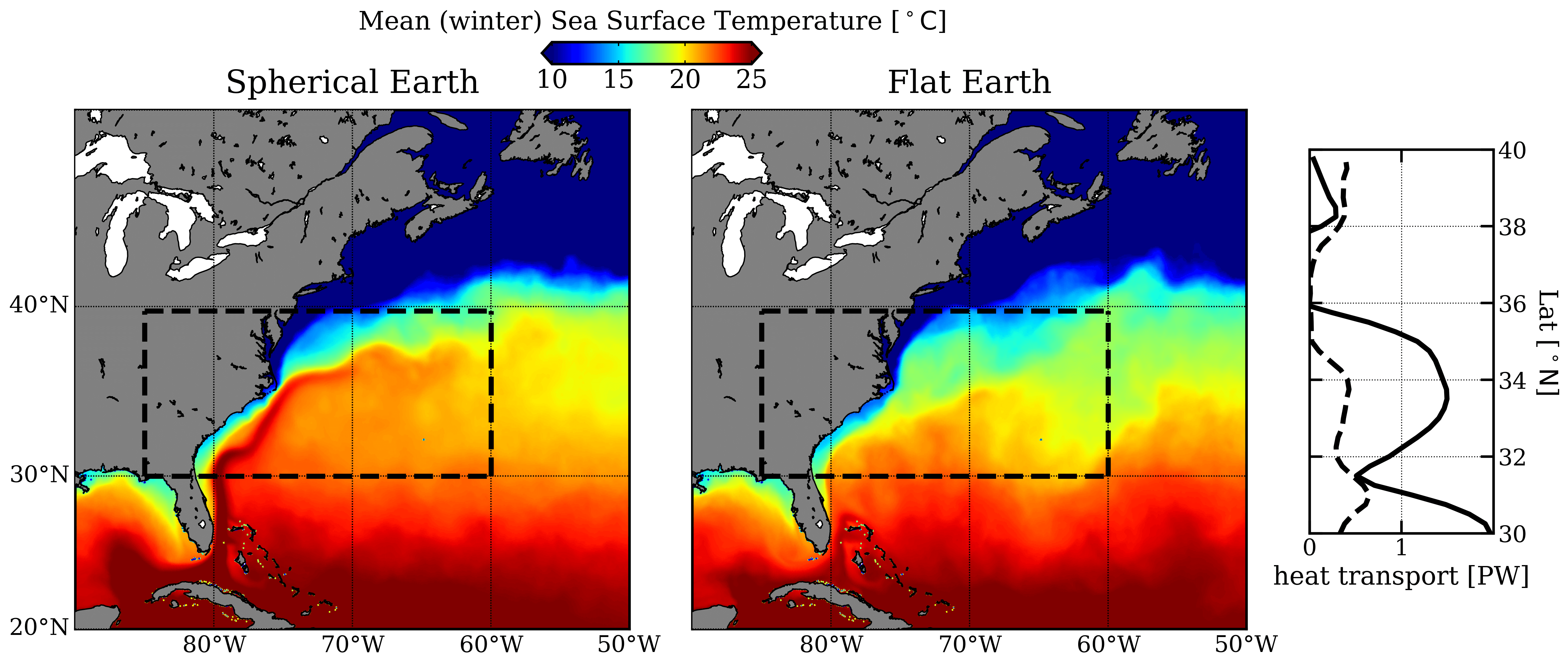}
\caption{(left) Winter averages of the SST in the BETA and the FLAT run; black dashed area indicates the area in which the heat transport is estimated. (right) Estimation of the heat transport induced by the Gulf Stream. Black solid lines corresponds to the BETA simulation while black dashed line corresponds to the FLAT simulation. To allow a finest comparison, simulation outputs are visualized with a Mercator projection.}
\label{sst}
\end{figure}

From these two simplistic comparisons, one concludes that Earth must not be flat.
If it were not, the basin-scale circulation of the North Atlantic would disappears, and so would the Gulf Stream. The loss of such an intense current would have impacts clearly visible on the North Atlantic climate. Indeed, the Gulf Stream carries warm tropical water northward, particularly during winter (see Fig. \ref{sst}). This creates a positive meridional heat transport \cite{bryden_heat_1980}. This latter phenomenon is important for the whole atmosphere dynamics \cite{minobe_influence_2008} and the European climate \cite{palter_role_2015}. In the FLAT simulation, we observe that if the earth were flat, the meridional heat transport in the region of the Gulf Stream, between 30 and 36$^\circ$N, would be divided by a factor of 8 (see Fig. \ref{sst}, right panel). As further evidence that the Earth is not flat, one can put forth that European as well as east U.S. climate would be significantly different (and temperature most likely cooler in winter) on a flat Earth.

\section{Discussion: why does the sphericity matter so much ?}

We have seen in a realistic context that the sphericity of the earth is of key importance for the large scale dynamics in the ocean, and particularly for the occurrence of western boundary currents such as the Gulf Stream. Note that similar observations could be made in other basins, such as the North Pacific, where the western boundary current occurring there is the Kuroshio \cite{mizuno_annual_1983}.
In this section, we use basic textbooks theories to illustrate why the sphericity of Earth matters for the basin scale dynamics.

The occurrence of the North Atlantic subtropical gyre can be explained using the \textit{Sverdrup relation}. As previously mentioned, the ocean mostly gains energy from the wind stress. Assuming that movements are slow and with a large horizontal extent ($T\gg1$ day and $L>100$ km), and considering a wind stress $\boldsymbol{\tau}=(\tau_x,\tau_y)$, the horizontal components of equations (\ref{prim}) become

\begin{equation}
  \left\{
      \begin{aligned}
        &-\rho_0fv=-\partial_x P+\partial_z\tau_x\\
        &+\rho_0fu=-\partial_y P+\partial_z\tau_y.\\
      \end{aligned}
    \right.
\label{tau_geo}
\end{equation}
If we vertically integrate these equations over the whole water column (considering a flat bottom), taking the curl of the result leads to

\begin{equation}
\partial_x(fU)+\partial_y(fV)=\boldsymbol{\nabla}\times\boldsymbol{\tau}|_z,
\label{tau_geo}
\end{equation}
where $\textbf{U}=(U,V)=\int dz \rho_0 \textbf{u}$ represents the barotropic component of the velocity (\textit{i.e} the part of the velocity which is homogeneous throughout the whole water column). Considering the $\beta$-plane approximation (\ref{beta}) and the non divergence of $\textbf{U}$ leads to the Sverdrup relation:

\begin{equation}
\beta V=\boldsymbol{\nabla}\times\boldsymbol{\tau}|_z.
\label{sverdrup}
\end{equation}
Because the winds are mainly zonal, this can be simplified as

\begin{equation}
\beta V=-\partial_y\tau_x.
\label{VV}
\end{equation}
Integrating the non-divergence $\partial_xU+\partial_yV=0$ zonally from west to east (with a reference $U=0$ at the western part of the basin) leads to an equation on the zonal velocity:

\begin{equation}
\beta U=(x-L)\,\partial_{yy}^2\tau_x,
\label{UU}
\end{equation}
with $x$ the meridional coordinate, and $L$ the meridional size of the basin. Notice that integrating from east to west gives a wrong result. This can be fixed by adding a bottom friction to equations (Stommel equations). 
%We kept here the simplest set of equations.

Above the North Atlantic, winds are from west to east in the northern part of the basin, and in the opposite direction near the Equator. Following equations (\ref{VV}) and (\ref{UU}) thus leads to a theoretical anticyclonic barotropic circulation at the basin-scale: the subtropical gyre. From this circulation results a positive anomaly of SSH. This is what we see for the spherical Earth simulation in Fig. \ref{ssh}. On a flat Earth, $\beta=0$, and equations (\ref{VV}) and (\ref{UU}) are no longer valid. The basin scale circulation is not set up, even if the wind blows the same way. This can be seen in the FLAT simulation outputs. 

In the primitive equations framework, a generalization of the Sverdrup relation (\ref{sverdrup}) can be obtained by integrating equations (\ref{prim}) ---in which we add all friction terms--- and cross-differentiating the result \cite{gula_gulf_2015}. This gives the so-called full barotropic vorticity balance equation:

\begin{equation}
\frac{\partial \Omega}{\partial t}=\overbrace{-\boldsymbol{\nabla}.(f\textbf{U})}^{\rm A} +  
\underbrace{\boldsymbol{\nabla}\times\boldsymbol{\tau}|_z}_{\rm B}+ 
\overbrace{\frac{\textbf{J}(P_b,h)}{\rho_0}}^{\rm C}+ 
\underbrace{\boldsymbol{\nabla}\times\boldsymbol{\tau}^{\rm bot}|_z}_{\rm D}+
\mathcal{D}_\Sigma+\mathcal{A}_\Sigma.
\label{full_vort}
\end{equation}
The left hand side of this relation is the temporal variation of the barotropic vorticity $\Omega=\partial_x V-\partial_y U$. The terms A and B are respectively the planetary vorticity advection and the wind curl. Because we consider long time averages, $\boldsymbol{\nabla}.(f\textbf{U})\approx \beta V$, and $\frac{\partial \Omega}{\partial t}\approx 0$. The balance between these two terms is the Sverdrup relation (\ref{sverdrup}). The term C is the bottom pressure torque with $\textbf{J}$ the Jacobian operator, $P_b$ the pressure anomaly along a contour line of fixed topography, and $h$ the topography. This term represents the local steering of oceanic currents by topographic features. The term D is the bottom drag curl. Finally, $\mathcal{D}_\Sigma$ and $\mathcal{A}_\Sigma$ are the horizontal diffusion term (induced by the horizontal advective scheme of the numerical model), and the nonlinear advection term respectively. Details about the last term, and how each term is computed from the model outputs can be found in \cite{gula_gulf_2015}.

\begin{figure}
\centering
\includegraphics[width=12cm]{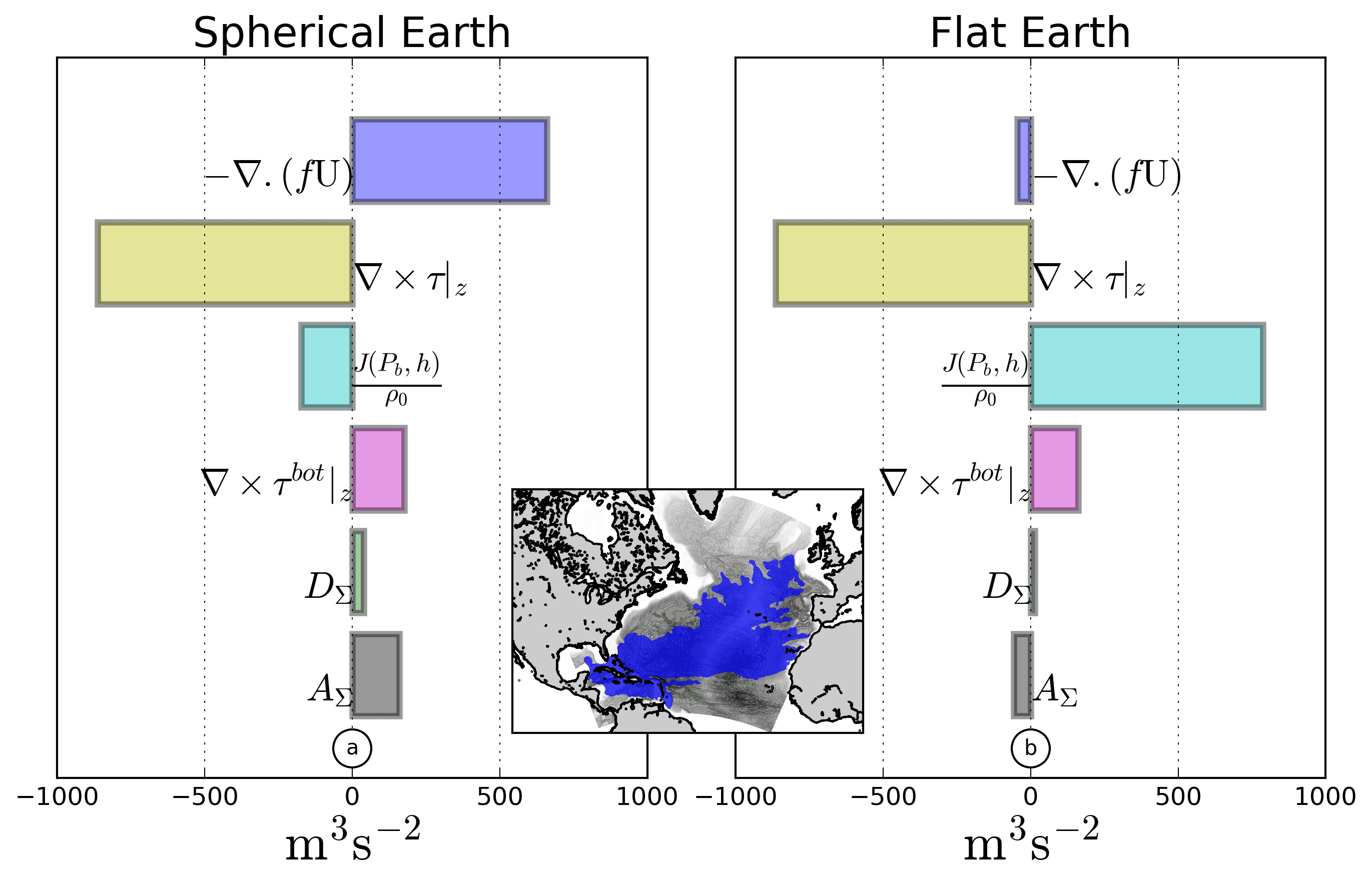}
\caption{Horizontally integrated terms of the barotropic vorticity balance equation for (left) the BETA simulation and (right) the FLAT simulation. The domain for the integration is shown by the blue contour in the middle insert.}
\label{vort}
\end{figure}

We horizontally integrate the time averages of the different terms of equation (\ref{full_vort}) for the BETA and the FLAT simulations (Fig. \ref{vort}). The integration domain corresponds to the subtropical gyre of the North Atlantic observed in the BETA simulation (a closed contour of streamfunction). The Gulf Stream is excluded from this domain, to only consider the Sverdrup-type circulation.

In the BETA simulation, we mainly observe a balance between the wind stress and the planetary vorticity, as suggested by the Sverdrup relation. In the FLAT simulation, because of the lack of $\beta$, the wind stress has to be compensated by other terms: here by the bottom pressure torque. The small topographic features are thus key for the basin scale circulation on a flat Earth. 
%This motivates the use of realistic simulations to understand what could have been the basin scale circulation on a flat Earth.

\textit{In summary}, the sphericity of the earth allows the basin scale circulation to set up, through a balance between the wind stress and the planetary vorticity induced by the existence of $\beta$.

As seen in the calculation of equation (\ref{UU}), an asymmetry appears in the ocean when considering the west to east direction or the east to west one. This is due to the propagation of \textit{Rossby waves}, which transport energy at large scales from east to west. To describe them theoretically, it is useful to simplify the primitive equations, to obtain the so-called \textit{quasi-geostrophic} model. It can be done by assuming a flow with $T\gg1$ day, $L>100$ km, and a small aspect ratio ($H/L\ll 1$, with $H$ the mean depth of the ocean). The development can be found in \cite{vallis2017atmospheric}. We consider here a one-layer ocean (an ocean which is not stratified), for simplicity, with no bottom topography, and the $\beta$-plane approximation. The governing equation reads:

\begin{equation}
\frac{d_g}{dt}\Big[ \Delta\Psi-\frac{\Psi}{R^2}+\beta y \Big]=0,
\label{qg}
\end{equation}
where $\Psi$ is the streamfunction (proportional to SSH, $\Psi=g\eta/f_0$), $R=\sqrt{gH/f_0^2}$ is the barotropic radius of deformation, and $\frac{d_g}{dt}$ is the material derivative considering only the geostrophic velocities. Linearizing (\ref{qg}) around a rest state, and introducing a perturbation of SSH as a monochromatic wave, gives the dispersion relation for Rossby waves:

\begin{equation}
\omega=\frac{-\beta k}{k^2+l^2+R^{-2}},
\label{disp}
\end{equation}
with $k$ and $l$ the zonal and meridional wavenumbers of the wave, and $\omega$ its pulsation. For large-scale motions ($k^2+l^2\ll R^{-2}$), this can be simplified as

\begin{equation}
\omega=-\beta k R^2,
\label{disp}
\end{equation}
which is the dispersion relation of long Rossby waves. The most salient feature of this equation is that the group velocity and the phase velocity of the wave are equal, $c=-\beta R^2$, and strictly negative. This means that the wave as well as its energy always travel toward the west. If the earth were flat, with $\beta=0$, Rossby waves are absent from the ocean. 

In the real ocean, the water column is stratified such that the water is heavier at the the bottom than at the surface. Thus, one have to take into account the vertical density variations in the expression of the dispersion relation of long Rossby waves. This can be done, in the simplest way, by replacing the barotropic radius of deformation $R$ by the first Rossby radius of deformation $R_d=NH/f$ that takes into account the local value of stratification $N$, depth $H$, and Coriolis parameter $f$. We refer the reader to \cite{chelton_geographical_1998} for a full description of $R_d$.

\begin{figure}
\centering
\includegraphics[width=16cm]{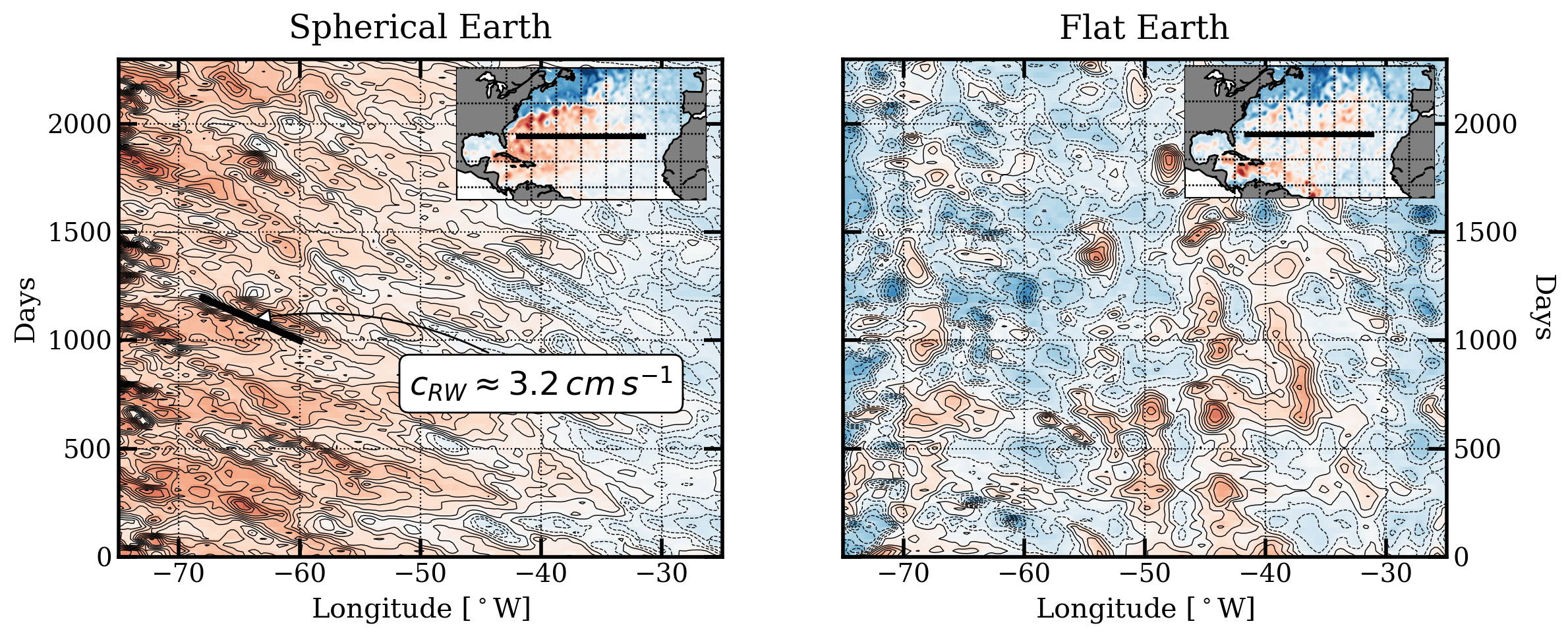}
\caption{Hovm\"{o}ller diagrams of the SSH along a line located at 30$^\circ$N (see inserts), for the BETA run (left) and the FLAT run (right). The black bold line in the left panel indicates the phase line of a theoretical long Rossby Wave with a phase velocity $c_{RW}=-\beta\,R_d^2$; $R_d$ is an estimation of the First Rossby Radius of deformation near 30$^\circ$N from \cite{chelton_geographical_1998}. }
\label{rw}
\end{figure}

Long Rossby waves propagating westward are seen in the BETA simulation. Indeed, the Hovm\"{o}ller diagram (time versus space diagram) of the SSH at the latitude of the Gulf Stream (30$^\circ$N), shows a streak pattern orientated westward (Fig. \ref{rw}). The slope of the streaks is in good accordance with the theoretical prediction of the phase speed of long Rossby waves. The time evolution of the SSH in the BETA simulation (not shown, see animation in supplementary materials) shows a westward movement of mesoscale structures. This movement is explained by the occurrence of long Rossby waves which advect eddies toward the west.
As the energy is carried toward the west by Rossby waves, physical processes are intensified at the western boundary of the basin. This explains why the North Atlantic gyre is intensified at the western boundary of the basin, forming the Gulf Stream.
In the FLAT simulation, no Rossby waves are seen. No Rossby wave can propagate because of the lack of $\beta$. The currents are not intensified at the western boundary, and mesoscale structures are nearly stationary. 

\textit{In summary}, the sphericity of the earth makes the energy travel towards the west. This explains the westward propagation of structures in the ocean as well as the western intensification of the oceanic currents.

\section{Conclusion}

With this study we have shown that a flat Earth cannot accurately reproduce the observed dynamics of the North Atlantic Ocean.
Indeed, on a flat Earth, numerous phenomena which have been intensively studied and observed in the past 70 years are not seen. Specific examples include the North Atlantic gyre, the Gulf Stream, the formation of mesoscale eddies, and their propagation toward the west. 

To be complete, it should be mentioned that the accuracy of the data and approximation of the governing equation of motions as implemented in the models may be questioned by supporters of the flat Earth. Here, we have focused on the Gulf Stream because its existence in the ocean is accepted by the Flat Earth Society. Explaining why the presence of the Gulf Stream requires the sphericity of the earth thus appeared to be a straightforward way toward the global \textit{re-}acceptation of the sphericity of the earth. 
%Thus, beside its pedagogical purpose, we believe that this study may help some people to understand that no \textit{internet forum} reasoning can explain why the earth should not be as it is. 

\section*{Acknowledgments}

We thank X. Carton for helpful discussions, and for his support in this unusual project. We also thank C. Buckingham for his precious comments that greatly improved the quality of the manuscript. Simulations were performed using the HPC facilities DATARMOR of “P\^ole de Calcul Intensif pour la Mer” at Ifremer, Brest, France.  Model outputs are available upon request, for people who are interested in using it, as well as for skeptics from all walks of life.

\bibliographystyle{unsrt}  
%\bibliographystyle{model5-names}
%\bibliography{bib_zotero}  

\end{document}